\documentclass[twocolumn,pra,amsmath,letterpaper,floatfix,showpacs]{revtex4}
\usepackage{graphicx}
\usepackage{amsmath}
\usepackage{amssymb}
\usepackage{color}

\begin{document}

\title{Magneto-optical properties of paramagnetic superrotors}

\author{A. A. Milner$^{1}$, A. Korobenko$^{1}$, J.~Flo{\ss}$^{2}$, I.Sh.~Averbukh$^{2}$, V. Milner$^{1}$}
\date{\today}

\affiliation{$^{1}$Department of  Physics \& Astronomy, The University of British Columbia, Vancouver, Canada \\
$^{2}$Department of Chemical Physics, The Weizmann Institute of Science, Rehovot, Israel}

\begin{abstract}
We study the dynamics of paramagnetic molecular superrotors in an external magnetic field. Optical centrifuge is used to create dense ensembles of oxygen molecules in ultra-high rotational states. It is shown for the first time, that the gas of rotating molecules becomes optically birefringent in the presence of magnetic field. The discovered effect of ``magneto-rotational birefringence'' indicates preferential alignment of molecular axes along the field direction. We provide an intuitive qualitative model, in which the influence of the applied magnetic field on the molecular orientation is mediated by the spin-rotation coupling. This model is supported by the direct imaging of the distribution of molecular axes, the demonstration of the magnetic reversal of the rotational Raman signal, and by numerical calculations.

\end{abstract}

\maketitle

Control of molecular rotation with electromagnetic fields of various types and geometries has become a powerful tool in a number of rapidly growing fields of molecular science (for a recent review, see Ref.\citenum{Lemeshko13}). From attosecond high-harmonic spectroscopy\cite{Kim14,Lepine14} and photoelectron spectroscopy\cite{Stolow08} to controlling molecular interactions with atoms\cite{Tilford04}, molecules\cite{Vattuone10} and surfaces\cite{Kuipers88, Zare98}, to altering molecular trajectories\cite{Purcell09, Gershnabel10}, the ability to align either the frames of rotating molecules or their angular momenta is one of the key requirements.

Sequences of linearly polarized ultrashort laser pulses have been successfully used for controlling molecular rotation\cite{Stapelfeldt03, Seideman05, Ohshima10, Fleischer12}. This excitation scheme, however, is limited to low rotational levels because of the detrimental effects of multi-photon ionization at high laser intensities. The rotation of polar or paramagnetic molecules can be harnessed by static electric or magnetic field, respectively \cite{Friedrich91, Friedrich92}. In both cases, an applied external field interacts with the permanent dipole moment of a molecule, hybridizing its rotational levels and creating pendular states in which the molecular axis is librating around the field direction\cite{Slenczka94}. The combination of static and optical fields has been suggested\cite{Friedrich99} and demonstrated\cite{Baumfalk01, Sakai03} as a means of enhancing molecular alignment beyond that achievable with a static field alone. All mentioned techniques rely on Stark or Zeeman shifts to exceed the energy separation of the rotational levels and are therefore limited to rotationally cold ensembles.

In this work, we extend the reach of rotational control by static external fields to ultra-high rotational states, which cannot be hybridized to form pendular states in laboratory magnetic fields. We use the method of molecular spinning in an optical centrifuge\cite{Villeneuve00, Yuan11} and exploit its capability to generate and control narrow rotational wave packets in extremely broad range of angular momenta\cite{Korobenko14a}. We show that a weak (on the rotational energy scale) external magnetic field of $<1$ T, applied at 90 degrees to the axis of molecular rotation, results in a significant re-orientation of molecular axes. We experimentally demonstrate this effect, hereafter referred to as ``magneto-rotational'', by observing that a gas of oxygen superrotors in an external magnetic field exhibits strong linear birefringence. A transparent physical mechanism of the observed phenomenon is proposed and confirmed by two complementary experimental techniques, and by numerical calculations. Set to ultrafast unidirectional rotation, centrifuged diatomic molecules (known as molecular superrotors) exhibit high degree of planar confinement as their atomic nuclei are strongly localized in the plane perpendicular to the axis of the centrifuge. For a superrotor with a nonzero magnetic moment coupled to its internuclear axis, such planar distribution is converted to the molecular alignment along the direction of the applied field and results in optical birefringence.

Magnetically induced birefringence is a powerful tool for establishing the spatial anisotropy of a wave function. In 1922, Gerlach used this effect in an (unsuccessful) attempt to detect the anisotropy of a disc-shaped electronic orbital in ground-state sodium atoms, expected from the Bohr model\cite{Estermann1975}. Applied to the excited electronic states, the same tool has been utilized in the pioneering work of Hanle on the polarization of resonant atomic fluorescence\cite{Hanle1924}, later extended to molecules\cite{German69}. The two key ingredients of the Hanle effect are the initial coherence between the non-degenerate electronic states (also evident from the closely related phenomenon of quantum beats in resonant emission\cite{Aleksandrov1973}) and a necessary mechanism of decoherence. While the coherent Larmor precession of the \textit{electronic angular momentum} makes the electronic wave function more isotropic, it is a competing relaxation process that results in the final anisotropic distribution, typically manifest in polarized fluorescence. The nature of the magneto-rotational birefringence, studied here, is rather different as neither the electronic coherence nor its decay are required. Instead, the precession of the \textit{rotational angular momentum} due to the spin-rotational interaction is solely responsible for the induced anisotropy of the molecular ensemble. Moreover, rather than being a result of decoherence, the effect is suppressed by any decoherence process.
\begin{figure}[tb]
\includegraphics[width=1\columnwidth]{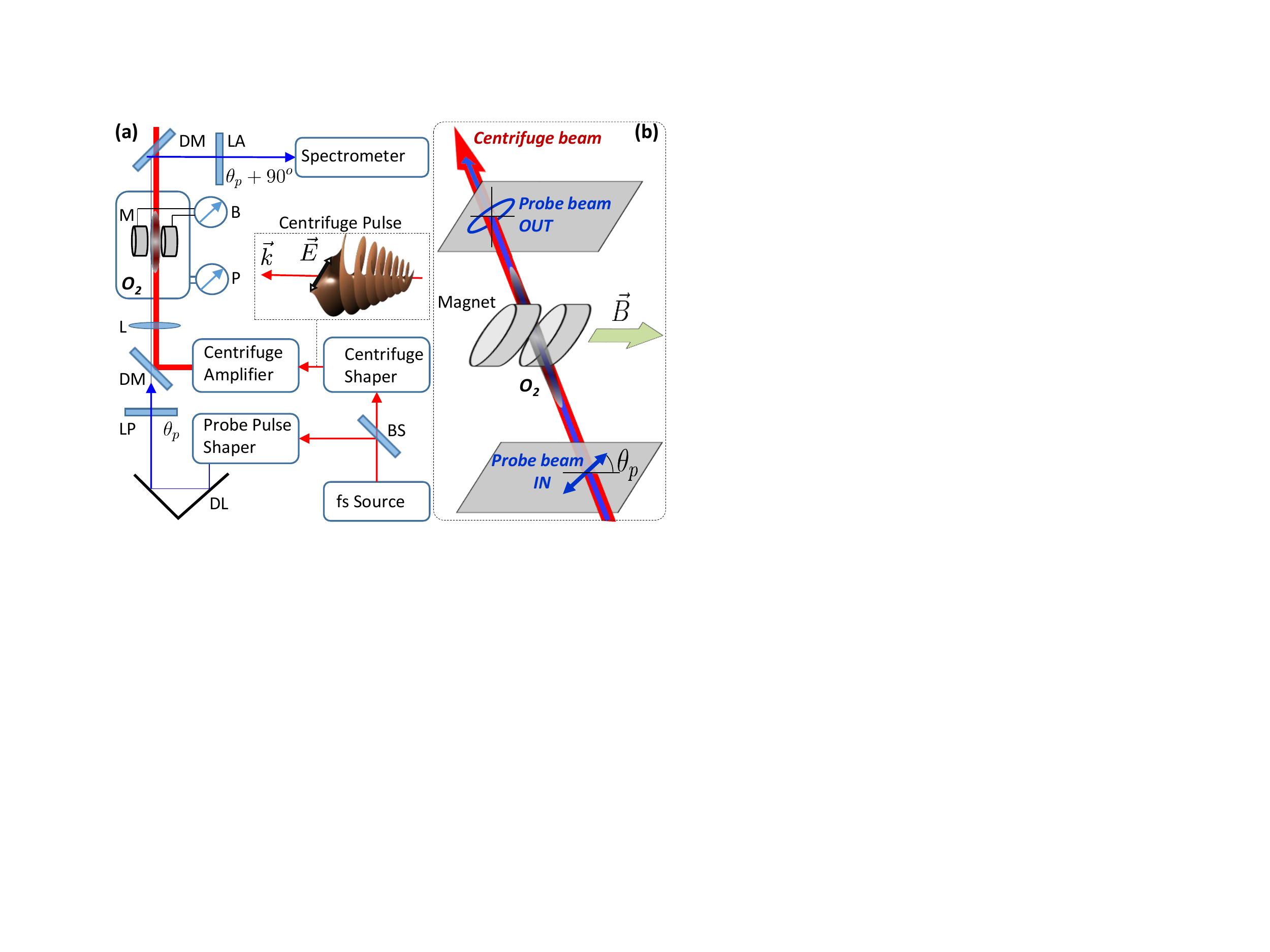}
\caption{(\textbf{a}) Experimental setup. BS: beam splitter, DM: dichroic mirror, LP (LA): linear polarizer (analyzer) oriented at angle $\theta_p$($\theta_p+90^{o}$) with respect to $\vec{B}$, DL: delay line, L: lens, M: two magnetic coils connected in a Helmholtz configuration. `O$_{2}$' marks the pressure chamber filled with oxygen gas under pressure $P$ at room temperature. An optical centrifuge field is illustrated above the centrifuge shaper with $\vec{k}$ being the propagation direction and $\vec{E}$ the vector of linear polarization undergoing an accelerated rotation. (\textbf{b}) Geometry of the magnetic and optical fields used in this work. The cloud of centrifuged molecules is depicted as a dark ellipse.}
\label{Fig-Setup}
\end{figure}

The experimental setup is shown in Fig.\ref{Fig-Setup}(\textbf{a}). A beam of femtosecond pulses from an ultrafast laser source (spectral full width at half maximum (FWHM) of 30 nm) is split in two parts. One part is sent to the ``centrifuge shaper'' which converts the input laser field into the field of an optical centrifuge (illustrated in the inset) according to the original recipe of Karczmarek \textit{et al.} \cite{Karczmarek99}. The centrifuge shaper is followed by a home built Ti:Sapphire multi-pass amplifier boosting the pulse energy to 50 mJ. The second (probe) beam is frequency shifted to 400 nm, spectrally narrowed to about 4 cm$^{-1}$ (FWHM) with a standard $4f$ Fourier pulse shaper and sent through the gas of centrifuged molecules. As demonstrated in our previous work on molecular superrotors\cite{Korobenko14a, Milner14a}, coherent molecular rotation results in the appearance of strong Raman sidebands in the output spectrum of probe pulses. Carried out with an $f$/4.8 spectrometer, state-resolved Raman detection  gives us an accurate way of measuring the frequency of molecular rotation, which is determined by the terminal angular frequency of the optical centrifuge.
\begin{figure}[b]
\includegraphics[width=1\columnwidth]{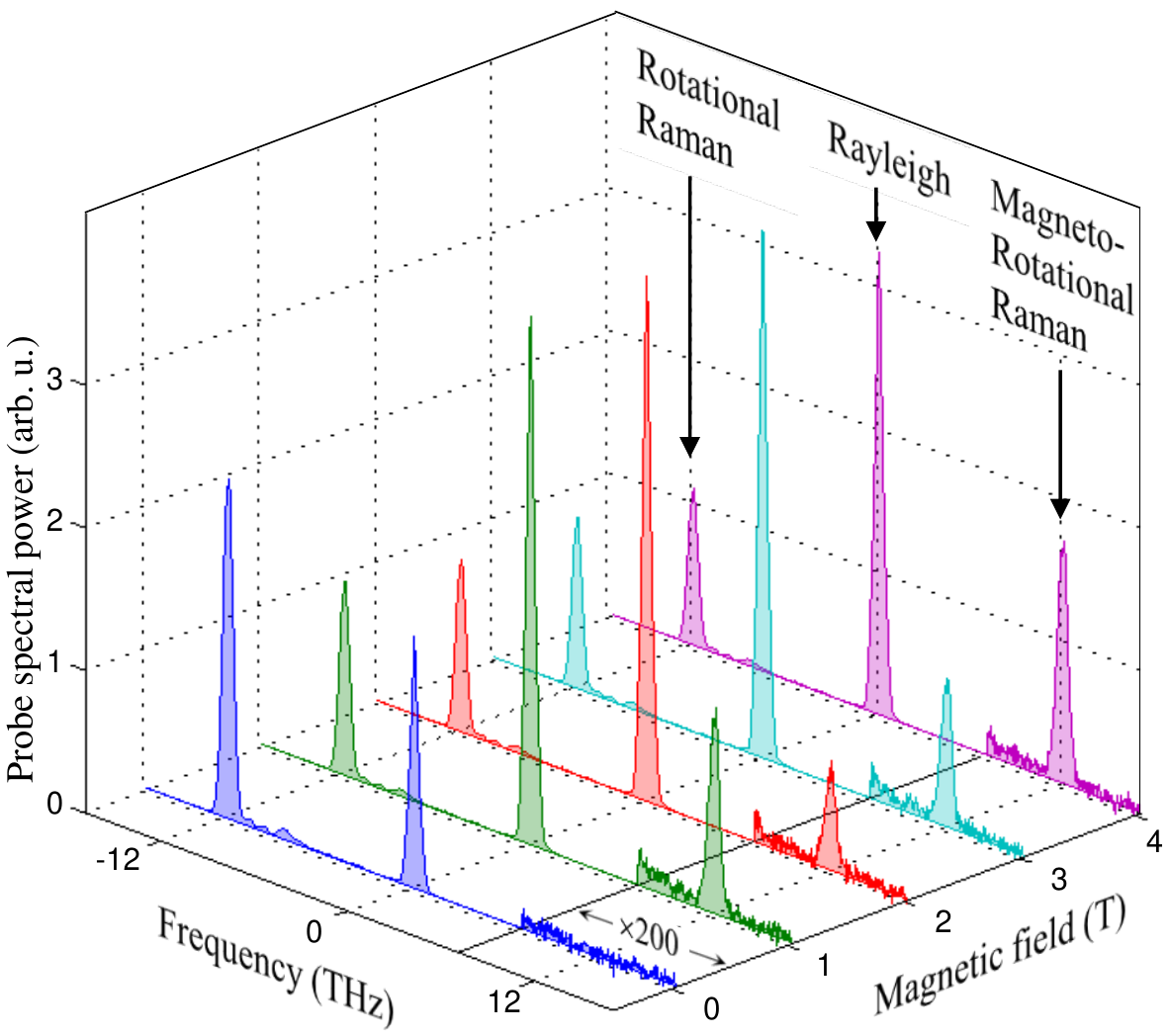}
\caption{Spectrum of probe pulses transmitted through the ensemble of centrifuged oxygen molecules as a function of the applied magnetic field. All spectra have been recorded at the probe delay of $t=1.14$ ns and under 0.3 atm of gas pressure. Crossed circular (rather than linear) polarizer and analyzer were used here to detect a weak magneto-rotational Raman signal (note the change of vertical scale ($\times 200$) at frequencies higher than 7 THz).}
\label{Fig-antiRaman}
\end{figure}

Both beams are focused collinearly between the two Helmholtz coils of a pulsed magnet inside the chamber filled with oxygen at room temperature and variable pressure (Fig.\ref{Fig-Setup}(\textbf{b})). The coils are driven by a 50 $\mu $s current pulse and produce magnetic fields of up to 4 T, which, on the time scale relevant to this study, can be considered static. After scattering off the centrifuged molecules, the polarization of the probe pulses becomes elliptical, reflecting linear optical birefringence of the molecular ensemble induced by the combination of the centrifuge and magnetic fields. To characterize the induced magneto-rotational birefringence, we pass probe pulses through a pair of crossed linear polarizers, LP and LA in Fig.\ref{Fig-Setup}(\textbf{a}).

In Fig.\ref{Fig-antiRaman}, the observed power spectrum of probe pulses, passed through the ensemble of centrifuged oxygen, is plotted as a function of the magnetic field strength. At zero field, the probe spectrum consists of two components: the frequency-unshifted Rayleigh line and the down-shifted (Stokes) Raman peak. The former is a result of the depolarization of probe light due to the non-uniform (and randomly changing from pulse to pulse) distribution of molecular axes in the plane of rotation. The second component reflects coherent molecular rotation with an angular frequency of about 6 THz (half the Raman shift). Oxygen molecules rotating with this frequency occupy rotational quantum states with $N=71$. The absence of the opposite anti-Stokes Raman line indicates unidirectional molecular rotation\cite{Korobenko14a}.

When the magnetic field is turned on, the amplitude of the Rayleigh peak becomes much higher, corresponding to strong incoherent (hence, frequency-unshifted) magneto-rotational birefringence. Unlike the weak initial depolarization which has no preferential axis at zero field, the magnetically induced birefringence exhibits a well defined optical axis. The applied magnetic field also affects the coherent Raman sidebands. While the initial Raman peak falls off with increasing $B$-field, a weak anti-Stokes Raman line grows on the other side of the spectrum, reflecting molecular rotation in the opposite, with respect to the centrifuge, direction.
\begin{figure}[t]
\includegraphics[width=1\columnwidth]{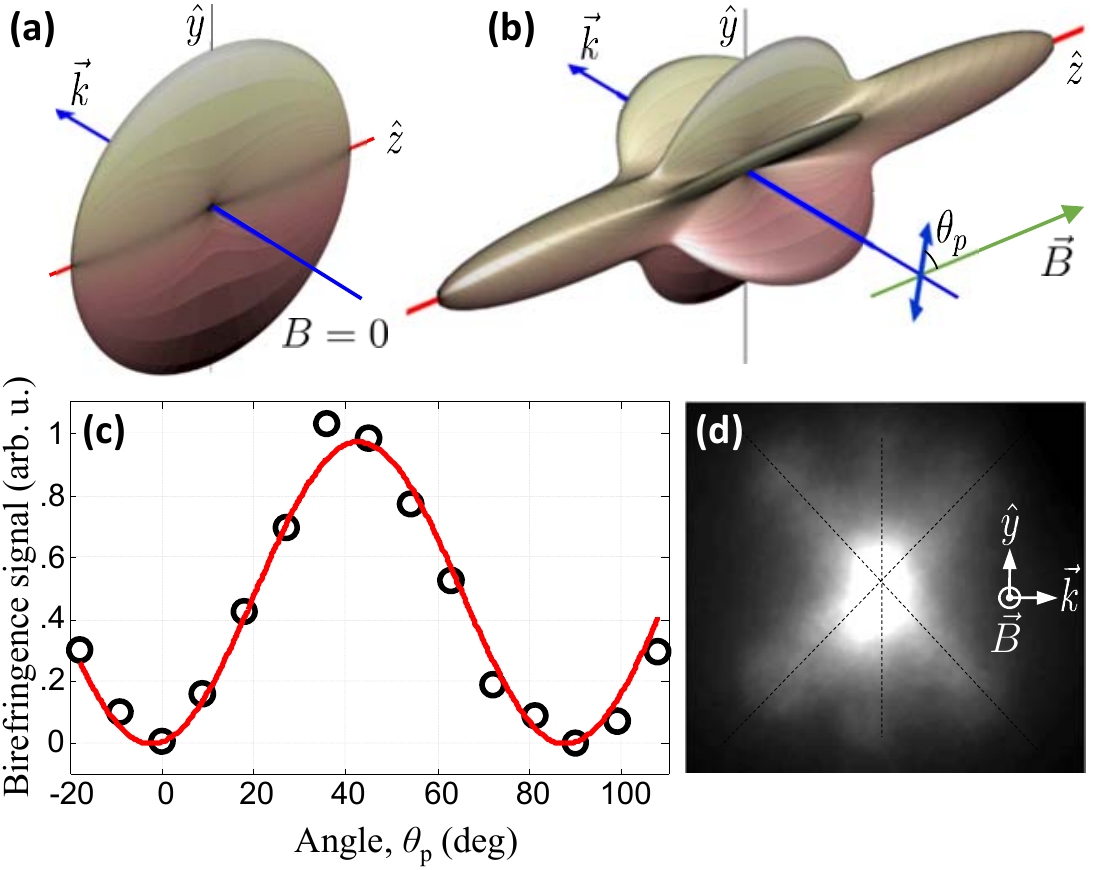}
\caption{(\textbf{a,b}) Calculated angular distribution of the molecular axes for the rotational state with $N=59$ at time $t=0.9$ ns in an external magnetic field of 0 and 0.32 Tesla, respectively. (\textbf{c}) Birefringence signal (scaled to peak at 1) as a function of angle $\theta_p$ between the polarization of probe pulses and the magnetic field direction. Black circles: data taken at 2 T, $t=1.5$ ns, and $N=95$. Red curve is a fit to $\cos^2(\theta_p)$. (\textbf{d}) Experimentally measured distribution of molecular axes, imaged in the direction of the applied field. All parameters are the same as in panel (\textbf{b}).}
\label{Fig-FieldAngle}
\end{figure}

To understand the nature of the observed magneto-rotational effects in the gas of paramagnetic superrotors, we calculate the angular distribution of O$_{2}$ molecules subject to constant magnetic field. The interaction between the spins of the two unpaired electrons and the nuclear rotation results in the spin-rotation coupling, which splits each rotational level in three components characterized by the total angular momentum $J=N, N\pm1$. An applied $B$-field lifts the degeneracy of each level with respect to the projection $M_{J}$ of $\vec{J}$ on the field axis. We calculate the energies $E_{N,J,M_J}(B)$ of the magnetic sublevels by numerically diagonalizing the electron-spin Zeeman Hamiltonian, starting from its matrix elements in Hund's case (b) basis set $|N,S,J,M_J \rangle$, with $S=1$ being the total electronic spin.

Given the initial thermal ensemble, the centrifuge creates an incoherent mixture of three states $|N,S,(J=N,N\pm1),M'_J=J \rangle$ with maximum projections $M'_J$ on the centrifuge direction. Using $J\gg S$, we approximate the angular distribution of the state $|N,S,J,M_J\rangle$ by the spherical harmonics $Y_{J,M_J}(\theta,\phi)$. We verified this approximation by a more elaborate calculation of the exact angular distributions which will be described in a future publication\cite{Floss2015}. Using this approximation, the final angular distribution at time $t$ is given as:
\begin{equation}\label{AngularDistribution}
    \rho_N (\theta,\phi) \approx \hspace{-3mm} \sum_{J=N, N\pm1}
    \Big| \sum_{M_J} c_{_{J,M_J}} e^{iE_{N,J,M_J}(B)t/\hbar} \, Y_{J,M_J}(\theta,\phi) \Big| ^2,
\nonumber
\end{equation}
where $c_{_{J,M_J}}$ are the wave function amplitudes in the coordinate frame with the quantization axis along the applied magnetic field. The results of our calculations for $N=59$ are shown in Fig.\ref{Fig-FieldAngle}. A disk-like distribution at $B=0$ (panel (\textbf{a})) corresponds to the molecular rotation around the centrifuge propagation direction $\vec{k}$, with the molecular axes isotropically distributed in the perpendicular $yz$ plane. As seen in panel (\textbf{b}) of Fig.\ref{Fig-FieldAngle}, an applied magnetic field splits the disk into three components. In the majority of cases considered here, the coupling between the electronic spin and an ultrafast rotation of superrotors is stronger than its interaction with the external B-field. The perturbative effect of the latter on the precession of the spin around $\vec{N}$ results in a clockwise (counter-clockwise) rotation of $\vec{N}$ around $\vec{B}$ for $J=N-1$ ($J=N+1$) states and no rotation for $J=N$ states. This leads to the anisotropic distribution stretched along $\hat{z}$. In agreement with this model, an experimentally measured $\cos^2(\theta_p)$ dependence shown in Fig.\ref{Fig-FieldAngle}(\textbf{c}) indicates the appearance of a well defined anisotropy axis along the direction of the field, as expected from the geometry of the calculated distribution. We note that for stronger magnetic fields, the three spin states decouple from the molecular angular momentum and the physical picture changes\cite{Floss2015}.

To validate the proposed model directly, we employ the technique of ion imaging for mapping out the distribution of molecular axes. Oxygen superrotors are Coulomb-exploded by a femtosecond probe pulse in a vacuum chamber, after traveling in an external magnetic field for a given amount of time\cite{Korobenko2015a}. In Fig.\ref{Fig-FieldAngle}(\textbf{d}), we show an image taken along the direction of the $B$-field ($\hat{z}$) under the same conditions as those used for the calculation in panel (\textbf{b}). A clear splitting into the three components supports the described mechanism behind the induced optical anisotropy.
\begin{figure}[t]
\includegraphics[width=1\columnwidth]{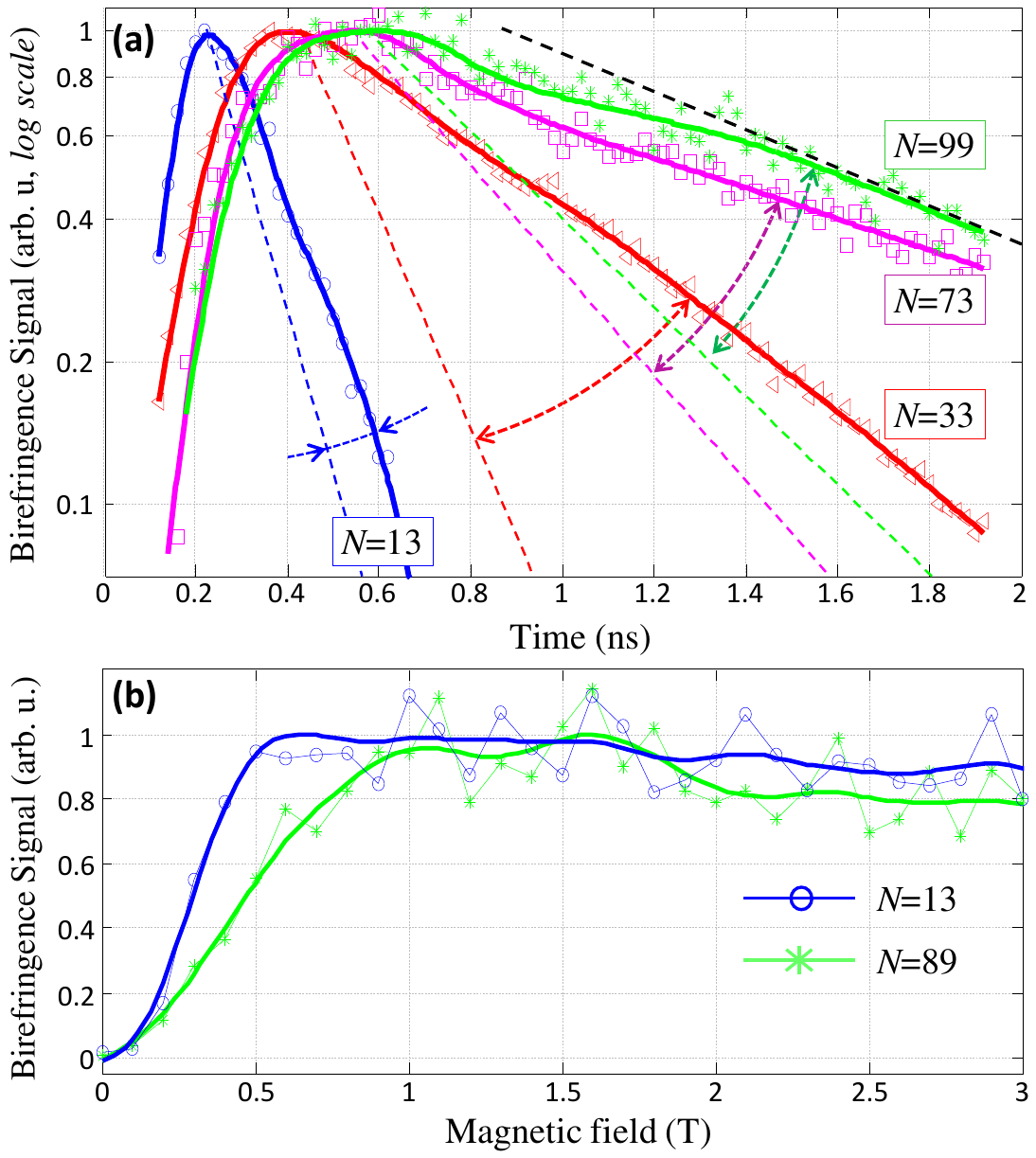}
\caption{(\textbf{a}) Decay of the birefringence signal for different values of the rotational quantum number at $B=2$ T.  All solid curves are generated by spline-fitting the experimental data (shown with colored markers) and normalized to peak at 1. The corresponding lifetimes are $85\pm10, 290\pm20, 660\pm50$ and $610\pm50$ ps$\cdot$atm for $N=13,33,73$ and 99, respectively. Colored dashed lines indicate the rate of rotational energy transfer due to the $N$-changing collisions\cite{Milner14b}. The black dashed line represents the asymptote of the rotational decoherence rate at the limit of ultra-high angular momenta. (\textbf{b}) Dependence of the birefringence signal recorded at $t=$1 ns on the strength of the applied magnetic field.}
\label{Fig-Dynamics}
\end{figure}

We now proceed with the analysis of the temporal evolution of the birefringence signal, which we define as the difference between the amplitude of the Rayleigh peak with and without the applied magnetic field. Figure \ref{Fig-Dynamics}(\textbf{a}) shows this signal, measured at a fixed magnetic field of 2 Tesla, as a function of the probe time delay with respect to the beginning of the centrifuge pulse. After a rapid initial growth, the magneto-rotational birefringence decays exponentially due to de-orienting collisions.

Truncating the centrifuge pulse in the centrifuge shaper so as to stop the accelerated rotation at different angular frequencies allows us to create rotational wave packets centered at different values of $N$\cite{Korobenko14a} and study the dependence of the magnetic effect on the frequency of molecular rotation. The results are plotted in Fig.\ref{Fig-Dynamics}(\textbf{a}) and indicate the decreasing decay rate of the magnetically induced birefringence with increasing $N$. One can also see that the birefringence signals reach their maximum values at different delay times: the higher the angular momentum, the slower the magnetic response.

To infer the mechanism of the observed decay, we compare it to the rate of rotational energy transfer due to collisions, analyzed in our recent study of oxygen superrotors\cite{Milner14b}. For each value of $N$, the rate of the $N$-changing collisions is indicated by a dashed line originating from the peak of the corresponding curve. The similarity of the two time scales points at the inelastic collisions as the main cause for the decay of the magneto-rotational birefringence. For all $N$'s except $N=13$, the birefringence signals decrease slower than the rotational population of the corresponding state. This fact suggests that the centrifuge-induced \textit{directionality} of molecular rotation is the reason behind the slow decay of the observed magnetic birefringence.  This is consistent with the recently found propensity of molecules to keep the orientation of their angular momentum in the course of inelastic collisions ("gyroscopic effect"), as reflected in the relaxation dynamics of the laser induced molecular alignment signals\cite{Hartmann12,Vieillard13}.

At low rotational frequencies ($N=13$), the gyroscope effect is weaker and the directionality of rotation becomes as susceptible to collisions as the rotational energy itself (blue curves in Fig.\ref{Fig-Dynamics}(\textbf{a})). The other end of the scale ($N\gtrapprox100$) corresponds to the limit of adiabatic collisions, where the rate of rotational decoherence approaches its asymptotic value and no longer depends on the rotational quantum number. Found in Ref.\citenum{Milner14b} and attributed to spin-flipping collisions, i.e. collisions which lead to transitions between the $J=N, N\pm1$ states within a single $N$ manifold, this decoherence rate is indicated by the black dashed line in the upper right corner of Fig.\ref{Fig-Dynamics}(\textbf{a}).

The dependence of the magneto-rotational birefringence on the strength of the applied magnetic field is shown in Fig.\ref{Fig-Dynamics}(\textbf{b}). After the initial growth with increasing $B$, the signal saturates. For different rotational frequencies, the saturation occurs at different field amplitudes. Slower superrotors do not only require shorter time to respond to the external magnetic field, but also need weaker fields to reach the plateau. Our numerical estimates of the birefringence signal, based on the calculated angular distributions of molecular axes (see Fig.\ref{Fig-FieldAngle}), successfully reproduce our experimental findings, showing qualitatively similar behavior.

Angular distributions are also instructive in the interpretation of the magnetic reversal of the coherent Raman scattering. The latter effect is clearly seen in Fig.\ref{Fig-antiRaman} as the appearance of a weak Raman line with a frequency shift corresponding to the molecular rotation in the direction opposite to the initial, centrifuge-induced rotation. Indeed, the precession of the two parts of the angular distribution around the applied magnetic field, as described earlier in the text, results in a small negative component in the projection of the molecular angular momentum on the centrifuge axis. The above mentioned effects of magnetic saturation and magnetic reversal will be further analyzed in the future publication.

To summarize, we showed that strong planar confinement ($\langle \cos \theta_{y,z} ^2 \rangle = 0.5$) of paramagnetic molecular superrotors can be converted into an anisotropic angular distribution ($\langle \cos \theta_y ^2 \rangle < 0.5, \langle \cos \theta_z ^2 \rangle = 0.5$) by means of external magnetic field. Here, $\theta_\alpha$ is the angle between the molecular axis and unit vector $\hat{\alpha}$, and $\langle .. \rangle$ represents ensemble averaging. Our experimental and theoretical analysis confirms the mediating role of an electronic spin in the demonstrated new scheme of rotation control. The effect can be used for fast switching of optical birefringence in gases, as well as for studying aligned molecules in ultra-high rotational states. Magneto-rotational birefringence, investigated in this work, also offers a way of exploring the effect of collisions on the directionality of molecular rotation in dense media.

We are grateful to Prof. E.~B.~Aleksandrov for reading our manuscript, providing us with an important review paper and offering his opinion on the topic of our work. We thank Prof. D.~R.~Herschbach for pointing out a historical connection to the work of Stern and Gerlach. This research has been supported by the grants from CFI, BCKDF, NSERC, ISF, DFG and the Minerva Foundation.

\begin{thebibliography}{34}
\expandafter\ifx\csname natexlab\endcsname\relax\def\natexlab#1{#1}\fi
\expandafter\ifx\csname bibnamefont\endcsname\relax
  \def\bibnamefont#1{#1}\fi
\expandafter\ifx\csname bibfnamefont\endcsname\relax
  \def\bibfnamefont#1{#1}\fi
\expandafter\ifx\csname citenamefont\endcsname\relax
  \def\citenamefont#1{#1}\fi
\expandafter\ifx\csname url\endcsname\relax
  \def\url#1{\texttt{#1}}\fi
\expandafter\ifx\csname urlprefix\endcsname\relax\def\urlprefix{URL }\fi
\providecommand{\bibinfo}[2]{#2}
\providecommand{\eprint}[2][]{\url{#2}}

\bibitem[{\citenamefont{Lemeshko et~al.}(2013)\citenamefont{Lemeshko, Krems,
  Doyle, and Kais}}]{Lemeshko13}
\bibinfo{author}{\bibfnamefont{M.}~\bibnamefont{Lemeshko}},
  \bibinfo{author}{\bibfnamefont{R.~V.} \bibnamefont{Krems}},
  \bibinfo{author}{\bibfnamefont{J.~M.} \bibnamefont{Doyle}}, \bibnamefont{and}
  \bibinfo{author}{\bibfnamefont{S.}~\bibnamefont{Kais}},
  \bibinfo{journal}{Mol. Phys.} \textbf{\bibinfo{volume}{111}},
  \bibinfo{pages}{1648} (\bibinfo{year}{2013}).

\bibitem[{\citenamefont{Kim et~al.}(2014)\citenamefont{Kim, Villeneuve, and
  Corkum}}]{Kim14}
\bibinfo{author}{\bibfnamefont{K.~T.} \bibnamefont{Kim}},
  \bibinfo{author}{\bibfnamefont{D.~M.} \bibnamefont{Villeneuve}},
  \bibnamefont{and} \bibinfo{author}{\bibfnamefont{P.~B.}
  \bibnamefont{Corkum}}, \bibinfo{journal}{Nat. Photonics}
  \textbf{\bibinfo{volume}{8}}, \bibinfo{pages}{187} (\bibinfo{year}{2014}).

\bibitem[{\citenamefont{Lepine et~al.}(2014)\citenamefont{Lepine, Ivanov, and
  Vrakking}}]{Lepine14}
\bibinfo{author}{\bibfnamefont{F.}~\bibnamefont{Lepine}},
  \bibinfo{author}{\bibfnamefont{M.~Y.} \bibnamefont{Ivanov}},
  \bibnamefont{and} \bibinfo{author}{\bibfnamefont{M.~J.~J.}
  \bibnamefont{Vrakking}}, \bibinfo{journal}{Nat. Photonics}
  \textbf{\bibinfo{volume}{8}}, \bibinfo{pages}{195} (\bibinfo{year}{2014}).

\bibitem[{\citenamefont{Stolow and Underwood}(2008)}]{Stolow08}
\bibinfo{author}{\bibfnamefont{A.}~\bibnamefont{Stolow}} \bibnamefont{and}
  \bibinfo{author}{\bibfnamefont{J.~G.} \bibnamefont{Underwood}}, in
  \emph{\bibinfo{booktitle}{Advances in Chemical Physics, Vol 139}}, edited by
  \bibinfo{editor}{\bibfnamefont{S.~A.} \bibnamefont{Rice}}
  (\bibinfo{publisher}{John Wiley \& Sons Inc}, \bibinfo{address}{New York},
  \bibinfo{year}{2008}), vol. \bibinfo{volume}{139} of
  \emph{\bibinfo{series}{Advances in Chemical Physics}}, pp.
  \bibinfo{pages}{497--583}.

\bibitem[{\citenamefont{Tilford et~al.}(2004)\citenamefont{Tilford, Hoster,
  Florian, and Forrey}}]{Tilford04}
\bibinfo{author}{\bibfnamefont{K.}~\bibnamefont{Tilford}},
  \bibinfo{author}{\bibfnamefont{M.}~\bibnamefont{Hoster}},
  \bibinfo{author}{\bibfnamefont{P.~M.} \bibnamefont{Florian}},
  \bibnamefont{and} \bibinfo{author}{\bibfnamefont{R.~C.}
  \bibnamefont{Forrey}}, \bibinfo{journal}{Phys. Rev. A}
  \textbf{\bibinfo{volume}{69}}, \bibinfo{pages}{052705}
  (\bibinfo{year}{2004}).

\bibitem[{\citenamefont{Vattuone et~al.}(2010)\citenamefont{Vattuone, Savio,
  Pirani, Cappelletti, Okada, and Rocca}}]{Vattuone10}
\bibinfo{author}{\bibfnamefont{L.}~\bibnamefont{Vattuone}},
  \bibinfo{author}{\bibfnamefont{L.}~\bibnamefont{Savio}},
  \bibinfo{author}{\bibfnamefont{F.}~\bibnamefont{Pirani}},
  \bibinfo{author}{\bibfnamefont{D.}~\bibnamefont{Cappelletti}},
  \bibinfo{author}{\bibfnamefont{M.}~\bibnamefont{Okada}}, \bibnamefont{and}
  \bibinfo{author}{\bibfnamefont{M.}~\bibnamefont{Rocca}},
  \bibinfo{journal}{Prog. Surf. Sci.} \textbf{\bibinfo{volume}{85}},
  \bibinfo{pages}{92} (\bibinfo{year}{2010}).

\bibitem[{\citenamefont{Kuipers et~al.}(1988)\citenamefont{Kuipers, Tenner,
  Kleyn, and Stolte}}]{Kuipers88}
\bibinfo{author}{\bibfnamefont{E.~W.} \bibnamefont{Kuipers}},
  \bibinfo{author}{\bibfnamefont{M.~G.} \bibnamefont{Tenner}},
  \bibinfo{author}{\bibfnamefont{A.~W.} \bibnamefont{Kleyn}}, \bibnamefont{and}
  \bibinfo{author}{\bibfnamefont{S.}~\bibnamefont{Stolte}},
  \bibinfo{journal}{Nature} \textbf{\bibinfo{volume}{334}},
  \bibinfo{pages}{420} (\bibinfo{year}{1988}).

\bibitem[{\citenamefont{Zare}(1998)}]{Zare98}
\bibinfo{author}{\bibfnamefont{R.~N.} \bibnamefont{Zare}},
  \bibinfo{journal}{Science} \textbf{\bibinfo{volume}{279}},
  \bibinfo{pages}{1875} (\bibinfo{year}{1998}).

\bibitem[{\citenamefont{Purcell and Barker}(2009)}]{Purcell09}
\bibinfo{author}{\bibfnamefont{S.~M.} \bibnamefont{Purcell}} \bibnamefont{and}
  \bibinfo{author}{\bibfnamefont{P.~F.} \bibnamefont{Barker}},
  \bibinfo{journal}{Phys. Rev. Lett.} \textbf{\bibinfo{volume}{103}},
  \bibinfo{pages}{153001} (\bibinfo{year}{2009}).

\bibitem[{\citenamefont{Gershnabel and Averbukh}(2010)}]{Gershnabel10}
\bibinfo{author}{\bibfnamefont{E.}~\bibnamefont{Gershnabel}} \bibnamefont{and}
  \bibinfo{author}{\bibfnamefont{I.~Sh.} \bibnamefont{Averbukh}},
  \bibinfo{journal}{Phys. Rev. Lett.} \textbf{\bibinfo{volume}{104}},
  \bibinfo{pages}{153001} (\bibinfo{year}{2010}).

\bibitem[{\citenamefont{Stapelfeldt and Seideman}(2003)}]{Stapelfeldt03}
\bibinfo{author}{\bibfnamefont{H.}~\bibnamefont{Stapelfeldt}} \bibnamefont{and}
  \bibinfo{author}{\bibfnamefont{T.}~\bibnamefont{Seideman}},
  \bibinfo{journal}{Rev. Mod. Phys.} \textbf{\bibinfo{volume}{75}},
  \bibinfo{pages}{543} (\bibinfo{year}{2003}).

\bibitem[{\citenamefont{Seideman et~al.}(2005)\citenamefont{Seideman, Hamilton,
  Berman, and Lin}}]{Seideman05}
\bibinfo{author}{\bibfnamefont{T.}~\bibnamefont{Seideman}},
  \bibinfo{author}{\bibfnamefont{E.}~\bibnamefont{Hamilton}},
  \bibinfo{author}{\bibfnamefont{P.~R.} \bibnamefont{Berman}},
  \bibnamefont{and} \bibinfo{author}{\bibfnamefont{C.~C.} \bibnamefont{Lin}},
  in \emph{\bibinfo{booktitle}{Advances In Atomic, Molecular, and Optical
  Physics}} (\bibinfo{publisher}{Academic Press}, \bibinfo{year}{2005}), vol.
  \bibinfo{volume}{Volume 52}, pp. \bibinfo{pages}{289--329}.

\bibitem[{\citenamefont{Ohshima and Hasegawa}(2010)}]{Ohshima10}
\bibinfo{author}{\bibfnamefont{Y.}~\bibnamefont{Ohshima}} \bibnamefont{and}
  \bibinfo{author}{\bibfnamefont{H.}~\bibnamefont{Hasegawa}},
  \bibinfo{journal}{Int. Rev. Phys. Chem.} \textbf{\bibinfo{volume}{29}},
  \bibinfo{pages}{619} (\bibinfo{year}{2010}).

\bibitem[{\citenamefont{Fleischer et~al.}(2012)\citenamefont{Fleischer,
  Khodorkovsky, Gershnabel, Prior, and Averbukh}}]{Fleischer12}
\bibinfo{author}{\bibfnamefont{S.}~\bibnamefont{Fleischer}},
  \bibinfo{author}{\bibfnamefont{Y.}~\bibnamefont{Khodorkovsky}},
  \bibinfo{author}{\bibfnamefont{E.}~\bibnamefont{Gershnabel}},
  \bibinfo{author}{\bibfnamefont{Y.}~\bibnamefont{Prior}}, \bibnamefont{and}
  \bibinfo{author}{\bibfnamefont{I.~Sh.} \bibnamefont{Averbukh}},
  \bibinfo{journal}{Isr. J. Chem.} \textbf{\bibinfo{volume}{52}},
  \bibinfo{pages}{414} (\bibinfo{year}{2012}).

\bibitem[{\citenamefont{Friedrich and Herschbach}(1991)}]{Friedrich91}
\bibinfo{author}{\bibfnamefont{B.}~\bibnamefont{Friedrich}} \bibnamefont{and}
  \bibinfo{author}{\bibfnamefont{D.~R.} \bibnamefont{Herschbach}},
  \bibinfo{journal}{Nature} \textbf{\bibinfo{volume}{353}},
  \bibinfo{pages}{412} (\bibinfo{year}{1991}).

\bibitem[{\citenamefont{Friedrich and Herschbach}(1992)}]{Friedrich92}
\bibinfo{author}{\bibfnamefont{B.}~\bibnamefont{Friedrich}} \bibnamefont{and}
  \bibinfo{author}{\bibfnamefont{D.~R.} \bibnamefont{Herschbach}},
  \bibinfo{journal}{Z. Phys. D} \textbf{\bibinfo{volume}{24}},
  \bibinfo{pages}{25} (\bibinfo{year}{1992}).

\bibitem[{\citenamefont{Slenczka et~al.}(1994)\citenamefont{Slenczka,
  Friedrich, and Herschbach}}]{Slenczka94}
\bibinfo{author}{\bibfnamefont{A.}~\bibnamefont{Slenczka}},
  \bibinfo{author}{\bibfnamefont{B.}~\bibnamefont{Friedrich}},
  \bibnamefont{and}
  \bibinfo{author}{\bibfnamefont{D.~R.}~\bibnamefont{Herschbach}},
  \bibinfo{journal}{Phys. Rev. Lett.} \textbf{\bibinfo{volume}{72}},
  \bibinfo{pages}{1806} (\bibinfo{year}{1994}).

\bibitem[{\citenamefont{Friedrich and Herschbach}(1999)}]{Friedrich99}
\bibinfo{author}{\bibfnamefont{B.}~\bibnamefont{Friedrich}} \bibnamefont{and}
  \bibinfo{author}{\bibfnamefont{D.~R.} \bibnamefont{Herschbach}}, 
  \bibinfo{journal}{J. Phys. Chem. A} \textbf{\bibinfo{volume}{103}}, \bibinfo{pages}{10280}
  (\bibinfo{year}{1999}).

\bibitem[{\citenamefont{Baumfalk et~al.}(2001)\citenamefont{Baumfalk, Nahler,
  and Buck}}]{Baumfalk01}
\bibinfo{author}{\bibfnamefont{R.}~\bibnamefont{Baumfalk}},
  \bibinfo{author}{\bibfnamefont{N.~H.} \bibnamefont{Nahler}},
  \bibnamefont{and} \bibinfo{author}{\bibfnamefont{U.}~\bibnamefont{Buck}},
  \bibinfo{journal}{J. Chem. Phys.} \textbf{\bibinfo{volume}{114}},
  \bibinfo{pages}{4755} (\bibinfo{year}{2001}).

\bibitem[{\citenamefont{Sakai et~al.}(2003)\citenamefont{Sakai, Minemoto,
  Nanjo, Tanji, and Suzuki}}]{Sakai03}
\bibinfo{author}{\bibfnamefont{H.}~\bibnamefont{Sakai}},
  \bibinfo{author}{\bibfnamefont{S.}~\bibnamefont{Minemoto}},
  \bibinfo{author}{\bibfnamefont{H.}~\bibnamefont{Nanjo}},
  \bibinfo{author}{\bibfnamefont{H.}~\bibnamefont{Tanji}}, \bibnamefont{and}
  \bibinfo{author}{\bibfnamefont{T.}~\bibnamefont{Suzuki}},
  \bibinfo{journal}{Phys. Rev. Lett.} \textbf{\bibinfo{volume}{90}},
  \bibinfo{pages}{083001} (\bibinfo{year}{2003}).

\bibitem[{\citenamefont{Villeneuve et~al.}(2000)\citenamefont{Villeneuve,
  Aseyev, Dietrich, Spanner, Ivanov, and Corkum}}]{Villeneuve00}
\bibinfo{author}{\bibfnamefont{D.~M.} \bibnamefont{Villeneuve}},
  \bibinfo{author}{\bibfnamefont{S.~A.} \bibnamefont{Aseyev}},
  \bibinfo{author}{\bibfnamefont{P.}~\bibnamefont{Dietrich}},
  \bibinfo{author}{\bibfnamefont{M.}~\bibnamefont{Spanner}},
  \bibinfo{author}{\bibfnamefont{M.~Y.} \bibnamefont{Ivanov}},
  \bibnamefont{and} \bibinfo{author}{\bibfnamefont{P.~B.}
  \bibnamefont{Corkum}}, \bibinfo{journal}{Phys. Rev. Lett.}
  \textbf{\bibinfo{volume}{85}}, \bibinfo{pages}{542} (\bibinfo{year}{2000}).

\bibitem[{\citenamefont{Yuan et~al.}(2011)\citenamefont{Yuan, Teitelbaum,
  Robinson, and Mullin}}]{Yuan11}
\bibinfo{author}{\bibfnamefont{L.}~\bibnamefont{Yuan}},
  \bibinfo{author}{\bibfnamefont{S.~W.} \bibnamefont{Teitelbaum}},
  \bibinfo{author}{\bibfnamefont{A.}~\bibnamefont{Robinson}}, \bibnamefont{and}
  \bibinfo{author}{\bibfnamefont{A.~S.} \bibnamefont{Mullin}},
  \bibinfo{journal}{PNAS} \textbf{\bibinfo{volume}{108}}, \bibinfo{pages}{6872}
  (\bibinfo{year}{2011}).

\bibitem[{\citenamefont{Korobenko et~al.}(2014)\citenamefont{Korobenko, Milner,
  and Milner}}]{Korobenko14a}
\bibinfo{author}{\bibfnamefont{A.}~\bibnamefont{Korobenko}},
  \bibinfo{author}{\bibfnamefont{A.~A.} \bibnamefont{Milner}},
  \bibnamefont{and} \bibinfo{author}{\bibfnamefont{V.}~\bibnamefont{Milner}},
  \bibinfo{journal}{Phys. Rev. Lett.} \textbf{\bibinfo{volume}{112}},
  \bibinfo{pages}{113004} (\bibinfo{year}{2014}).

\bibitem[{\citenamefont{Estermann and Foner}(1975)}]{Estermann1975}
\bibinfo{author}{\bibfnamefont{I.}~\bibnamefont{Estermann}} \bibnamefont{and}
  \bibinfo{author}{\bibfnamefont{S.~N.} \bibnamefont{Foner}},
  \bibinfo{journal}{Am. J. Phys.} \textbf{\bibinfo{volume}{43}},
  \bibinfo{pages}{661} (\bibinfo{year}{1975}).

\bibitem[{\citenamefont{Hanle}(1924)}]{Hanle1924}
\bibinfo{author}{\bibfnamefont{W.}~\bibnamefont{Hanle}}, \bibinfo{journal}{Z.
  Physik} \textbf{\bibinfo{volume}{30}}, \bibinfo{pages}{93}
  (\bibinfo{year}{1924}).

\bibitem[{\citenamefont{German and Zare}(1969)}]{German69}
\bibinfo{author}{\bibfnamefont{K.~R.} \bibnamefont{German}} \bibnamefont{and}
  \bibinfo{author}{\bibfnamefont{R.~N.} \bibnamefont{Zare}},
  \bibinfo{journal}{Phys. Rev.} \textbf{\bibinfo{volume}{186}},
  \bibinfo{pages}{9} (\bibinfo{year}{1969}).

\bibitem[{\citenamefont{Aleksandrov}(1973)}]{Aleksandrov1973}
\bibinfo{author}{\bibfnamefont{E.~B.} \bibnamefont{Aleksandrov}},
  \bibinfo{journal}{Phys. Usp.} \textbf{\bibinfo{volume}{15}},
  \bibinfo{pages}{436} (\bibinfo{year}{1973}).

\bibitem[{\citenamefont{Karczmarek et~al.}(1999)\citenamefont{Karczmarek,
  Wright, Corkum, and Ivanov}}]{Karczmarek99}
\bibinfo{author}{\bibfnamefont{J.}~\bibnamefont{Karczmarek}},
  \bibinfo{author}{\bibfnamefont{J.}~\bibnamefont{Wright}},
  \bibinfo{author}{\bibfnamefont{P.}~\bibnamefont{Corkum}}, \bibnamefont{and}
  \bibinfo{author}{\bibfnamefont{M.}~\bibnamefont{Ivanov}},
  \bibinfo{journal}{Phys. Rev. Lett.} \textbf{\bibinfo{volume}{82}},
  \bibinfo{pages}{3420} (\bibinfo{year}{1999}).

\bibitem[{\citenamefont{Milner et~al.}(2014{\natexlab{a}})\citenamefont{Milner,
  Korobenko, Hepburn, and Milner}}]{Milner14a}
\bibinfo{author}{\bibfnamefont{A.~A.} \bibnamefont{Milner}},
  \bibinfo{author}{\bibfnamefont{A.}~\bibnamefont{Korobenko}},
  \bibinfo{author}{\bibfnamefont{J.~W.} \bibnamefont{Hepburn}},
  \bibnamefont{and} \bibinfo{author}{\bibfnamefont{V.}~\bibnamefont{Milner}},
  \bibinfo{journal}{Phys. Rev. Lett.} \textbf{\bibinfo{volume}{113}},
  \bibinfo{pages}{043005} (\bibinfo{year}{2014}{\natexlab{a}}).

\bibitem[{\citenamefont{Flo{\ss}}(2015)}]{Floss2015}
\bibinfo{author}{\bibfnamefont{J.}~\bibnamefont{Flo{\ss}}},
  \bibinfo{journal}{arXiv:1504.03278}  (\bibinfo{year}{2015}).

\bibitem[{\citenamefont{Korobenko and Milner}(2015)}]{Korobenko2015a}
\bibinfo{author}{\bibfnamefont{A.}~\bibnamefont{Korobenko}} \bibnamefont{and}
  \bibinfo{author}{\bibfnamefont{V.}~\bibnamefont{Milner}},
  \bibinfo{journal}{arXiv:1504.01432}  (\bibinfo{year}{2015}).

\bibitem[{\citenamefont{Milner et~al.}(2014{\natexlab{b}})\citenamefont{Milner,
  Korobenko, and Milner}}]{Milner14b}
\bibinfo{author}{\bibfnamefont{A.~A.} \bibnamefont{Milner}},
  \bibinfo{author}{\bibfnamefont{A.}~\bibnamefont{Korobenko}},
  \bibnamefont{and} \bibinfo{author}{\bibfnamefont{V.}~\bibnamefont{Milner}},
  \bibinfo{journal}{New J. Phys.} \textbf{\bibinfo{volume}{16}},
  \bibinfo{pages}{093038} (\bibinfo{year}{2014}{\natexlab{b}}).

\bibitem[{\citenamefont{Hartmann and Boulet}(2012)}]{Hartmann12}
\bibinfo{author}{\bibfnamefont{J.-M.} \bibnamefont{Hartmann}} \bibnamefont{and}
  \bibinfo{author}{\bibfnamefont{C.}~\bibnamefont{Boulet}},
  \bibinfo{journal}{J. Chem. Phys.} \textbf{\bibinfo{volume}{136}},
  (\bibinfo{year}{2012}).

\bibitem[{\citenamefont{Vieillard et~al.}(2013)\citenamefont{Vieillard,
  Chaussard, Billard, Sugny, Faucher, Ivanov, Hartmann, Boulet, and
  Lavorel}}]{Vieillard13}
\bibinfo{author}{\bibfnamefont{T.}~\bibnamefont{Vieillard}},
  \bibinfo{author}{\bibfnamefont{F.}~\bibnamefont{Chaussard}},
  \bibinfo{author}{\bibfnamefont{F.}~\bibnamefont{Billard}},
  \bibinfo{author}{\bibfnamefont{D.}~\bibnamefont{Sugny}},
  \bibinfo{author}{\bibfnamefont{O.}~\bibnamefont{Faucher}},
  \bibinfo{author}{\bibfnamefont{S.}~\bibnamefont{Ivanov}},
  \bibinfo{author}{\bibfnamefont{J.~M.} \bibnamefont{Hartmann}},
  \bibinfo{author}{\bibfnamefont{C.}~\bibnamefont{Boulet}}, \bibnamefont{and}
  \bibinfo{author}{\bibfnamefont{B.}~\bibnamefont{Lavorel}},
  \bibinfo{journal}{Phys. Rev. A} \textbf{\bibinfo{volume}{87}},
  \bibinfo{pages}{023409} (\bibinfo{year}{2013}).

\end{thebibliography}

\end{document}